\documentclass[american]{svproc}

\makeatletter

\usepackage[T1]{fontenc}
\usepackage[latin9]{inputenc}
\usepackage{wrapfig}
\usepackage{amsmath}
\usepackage{amssymb}
\usepackage{graphicx}
\usepackage[numbers]{natbib}
\usepackage[all]{xy}
\usepackage{listings}
\usepackage{babel}
\usepackage{subfig}
\usepackage[all]{xy}
\usepackage{url}
\usepackage[hidelinks]{hyperref}
\usepackage[ruled,linesnumbered]{algorithm2e}

\newenvironment{myAlgorithm}[1]{
  \begin{algorithm}
  \DontPrintSemicolon
  \caption{#1}
}{
  \end{algorithm}
}
\SetKwProg{myCodeBlock}{}{}{}
\newcommand{\algoIndent}[1]{\myCodeBlock{}{#1}}
\let\oldnl\nl
\newcommand{\nolinenumber}{\renewcommand{\nl}{\let\nl\oldnl}}

\makeatletter \newcommand{\xyR}[1]{\makeatletter\xydef@\xymatrixrowsep@{#1}\makeatother}
\makeatletter \newcommand{\xyC}[1]{\makeatletter\xydef@\xymatrixcolsep@{#1}\makeatother}

\@ifundefined{showcaptionsetup}{}{%
 \PassOptionsToPackage{caption=false}{subfig}}
\makeatother

\newcommand{\nats}{\mathbb{N}}
\newtheorem{myclaim}{Claim}

\begin{document}

\title{A Random Network Model for the Analysis of Blockchain Designs
  with Communication Delay}


\author{Carlos Pinzón \and Camilo Rocha \and Jorge Finke}

\institute{Pontificia Universidad Javeriana, Cali} \maketitle

\begin{abstract}
  This paper proposes a random network model for blockchains, a
  distributed hierarchical data structure of blocks that has found
  several applications in various industries.  The model is
  parametric on two probability distribution functions governing block
  production and communication delay, which are key to capture the
  complexity of the mechanism used to synchronize the many distributed
  local copies of a blockchain.  The proposed model is equipped with
  simulation algorithms for both bounded and unbounded number of
  distributed copies of the blockchain. They are used to study
  \textit{fast} blockchain systems, i.e., blockchains in which the
  average time of block production can match the average time of
  message broadcasting used for blockchain synchronization. In
  particular, the model and the algorithms are useful to understand
  \textit{efficiency} criteria associated with fast blockchains for
  identifying, e.g., when increasing the block production will have
  negative impact on the stability of the distributed data structure
  given the network's broadcast delay.
\end{abstract}

\section{Introduction}
\label{sec.intro}

A blockchain is a distributed hierarchical data structure of blocks
that cannot be altered retroactively without alteration of all
subsequent blocks, which requires consensus of the network majority.
It was invented to serve as the public transaction ledger of the
cryptocurrency Bitcoin in 2008~\citep{nakamoto2008bitcoin}, in which
the need for a trusted third party is avoided: instead, this digital
currency is based on the concept of `proof of work' allowing users to
execute payments by digitally signing their transactions using hashes
through a distributed time-stamping service~\citep{pinzon2016double}.
Because of its resistance to modifications, decentralized consensus,
and proved robustness for supporting cryptocurrency transactions,
this technology is seen to have great potential for new uses in other
domains, including financial
services~\citep{guo2016blockchain,tapscott2017blockchain}, distributed
data models~\citep{bui2018application},
markets~\citep{sikorski2017blockchain}, government
systems~\citep{hou2017application,olnes2017blockchain},
healthcare~\citep{gutierrez2018bc,alhadhrami2017introducing,karafiloski2017blockchain},
IoT~\citep{huh2017managing}, and video
games~\citep{munir2019challenges}. As the blockchain technology
matures, it is expected to change economics, business, and
society~\citep{aste2017blockchain} in the years to come.


Technically, a blockchain is a distributed append-only data structure
comprising a linear collection of blocks, shared among several
\textit{workers} (or, \textit{miners}; i.e., computational nodes
responsible for working on the blockchain with the goal to extend it
further with new blocks). Since the blockchain is decentralized, each
worker possesses a local copy of the blockchain. This means, e.g., that
workers can be working at the same time on unsynchronized local copies
of the blockchain structure. In the typical peer-to-peer network
implementation of blockchain technology, workers adhere to a consensus
protocol for inter-node communication and validation of new blocks: to
work on top of the largest blockchain and, in case of ties, on the one
whose last produced block was seen earliest.  This protocol guarantees
an effective synchronization mechanism as long as the task of
producing new blocks is hard to achieve, which is known in the
literature as `proof of work' blockchains, in comparison to the time
it takes for inter-node communication. The idea is that, if several
workers extend different versions of the blockchain, the consensus
mechanism allows the network to eventually select only one of them,
while the other blocks are discarded (including the data they carry)
when local copies are synchronized. This process carries on upon the
creation of new blocks.

The scenario of discarding blocks massively, which can be seen as an
efficiency issue in a blockchain implementation, is rarely present in
``slow'' block-producing blockchains. This is because the time it
takes to produce a new block in such implementations is long enough
for workers to synchronize their local copy of the blockchain. This
prevents, up to some extent, workers from wasting resources and time
in producing blocks that will likely be discharged during a future
synchronization stage. This is the case, e.g., of the Bitcoin network
in which it takes, in average, 10 minutes to mine a block and 12.6
seconds to communicate~\citep{decker2013information}; it was estimated
that the theoretical fork-rate of its blockchain was approximately
1.78\% in 2013~\citep{decker2013information}.  However, as the
blockchain technology finds new uses, it is being argued that block
production needs to be faster in order to have more versatile and
attractive applications~\citep{chohan2019limits,croman2016scaling}.
This means that precisely understanding how block production speed-ups
can negatively impact blockchains in terms of the amount of blocks
discharged due to race conditions among the workers is of great
practical importance for designing efficient blockchains in the near
future.

This paper presents a random network model for blockchains. It is
parametric on the number of workers under consideration (possibly
infinite), a probability distribution describing the time for
producing new blocks, and a probability distribution describing the
communication delay between any pair of random workers in the network.
The model is equipped with probabilistic algorithms that can be used
to mathematically simulate and analyze blockchains having a fixed or
unbounded number of workers producing blocks at different rates over a
network with communication delays. In this paper, the model and the
algorithms are used as means to study the continuous process of block
production in \textit{fast} blockchains: highly distributed networks
of workers rapidly producing blocks and in which inter-node
communication delays can be crucial for efficient block production. As
explained above, one of the main consequences of having faster block
production is that blocks tend to be discharged at a higher rate,
yielding a speed-efficiency tradeoff in fast blockchains. In this
work, experiments are presented to understand how such a tradeoff can
be analyzed for many scenarios to showcase the proposed approach. In
the broader picture of the years to come in which the use of fast
blockchain systems is likely to spread, the model, algorithms, and
approach contributed in this work can be seen as useful mathematical
tools for specifying, simulating, and analyzing such designs.


This paper is organized as follows. Section~\ref{sec.pow} summarizes
basic notions of proof-of-work blockchains. Sections~\ref{sec.model}
and~\ref{sec.algo} introduce the proposed network model and simulation
algorithms. Section~\ref{sec.results} presents experimental results in
the analysis of fast blockchains. Section~\ref{sec.concl} summarizes
related work and concludes the paper.

\section{An Overview of Proof-of-work Blockchains}
\label{sec.pow}

This section presents an overview of proof-of-work distributed
blockchain systems, including basic definitions and an example.

A \textit{blockchain} is a distributed hierarchical data structure of
blocks that cannot be altered retroactively without alteration of all
subsequent blocks, which requires consensus of the network majority.
The nodes in the network, called \emph{workers}, use their
computational power to generate \emph{blocks} with the goal of
extending the blockchain. The adjective `proof-of-work' comes from the
fact that producing a single block for the blockchain tends to be a
hard computational task for the workers (i.e., one requiring extensive
computation, e.g., a partial hash inversion).

\begin{definition}\label{def.pow.block}
  A \emph{block} is a digital document containing: (i) a digital
  signature stating the worker who produced it; (ii) an easy to verify
  proof-of-work witness in the form of a nonce; and (iii) a hash
  pointer to the previous block in the sequence (except for the first
  block, called the \emph{origin}, that has no previous block and is
  unique).
\end{definition}

A technical definition of blockchain as a data structure has been
proposed by different authors (see,
e.g.,~\citep{10.3389/fbloc.2019.00003}), although most of them
coincide on it being an immutable, transparent, and decentralized data
structure shared by all the workers in the network. For the purpose of
this paper, it is important to clearly distinguish between the
\textit{local} copy of the blockchain each worker has and the abstract
\textit{global} blockchain shared by all workers. It is the latter one
that holds the complete history of the blockchain.

\begin{definition}\label{def.pow.blockchain}
  The \emph{local blockchain} of a worker $w$ is a non-empty sequence
  of blocks stored in the local memory of $w$. The \emph{global
    blockchain} (or, \emph{blockchain}) is the minimal rooted tree
  containing all workers' local blockchains as branches.
\end{definition}

Note that by the assumption that the root node is unique
(Definition~\ref{def.pow.block}), the (global) blockchain is
well-defined no matter how many workers are part of the network and is
non-empty if there is at least one
worker. Definition~\ref{def.pow.blockchain} allows for local
blockchains of workers to be either synchronized or unsynchronized,
the latter being more common in systems with long propagation times or
in the presence anomalous situations (e.g., if there is an attacker
intentionally trying to hold a fork). This is a good reason why the
global blockchain cannot simply be defined as a unique sequence of
blocks, but rather as a distributed data structure to which workers
can be assumed to be partly synchronized.

\begin{figure}
  \centering
$\xymatrix{\xyR{0pc}\xyC{0.75pc}w_{0}:0 & 1\ar[l] & 5\ar[l] &  & 0 & 1\ar[l] & 5^{w_{0}}\ar[l]\\
w_{1}:0 & 2\ar[l] & 3\ar[l] & 6\ar[l] &  & 2\ar[ul] & 3^{w_{3}}\ar[l] & 6^{w_{1},w_{4}}\ar[l]\\
w_{2}:0 & 2\ar[l] & 4\ar[l] &  &  &  & 4^{w_{2}}\ar[ul]\\
w_{3}:0 & 2\ar[l] & 3\ar[l]\\
w_{4}:0 & 2\ar[l] & 3\ar[l] & 6\ar[l]
}
$
\caption{\label{fig.pow.blochainrep}A blockchain network of five
  workers with their local blockchains (left) and the corresponding
  global blockchain (right); blocks are represented by natural
  numbers.  Workers $w_{0}$, $w_{2}$, and $w_{3}$ are not yet
  synchronized with the longest sequence of blocks.}
\end{figure}
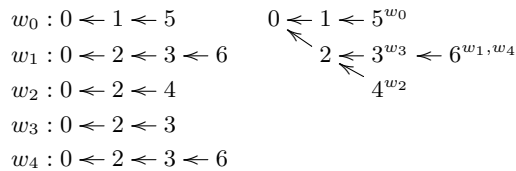

Figure~\ref{fig.pow.blochainrep} presents an example of a blockchain
with five workers, where blocks are represented by natural numbers. On
the left, the local blockchains are depicted as linked lists; on the
right, the corresponding global blockchain is depicted as a rooted
tree. Some of the blocks in the rooted tree representation in
Figure~\ref{fig.pow.blochainrep} are labeled with the identifier of a
worker: it indicates the position of each worker in the global
blockchain.  For modeling purposes, the rooted tree representation of
a blockchain is preferred. This is because, on the one hand, it can
reduce the amount of memory needed to store it and, on the other hand,
it visually simplifies the structure analysis of the data structure.
More specifically, storing a global blockchain with $m$ workers
containing $n$ unique blocks as a collection of lists requires $O(mn)$
memory in the worst case (i.e., with perfect synchronization).  In
contrast, the rooted tree representation of the same blockchain with
$m$ workers and $n$ unique blocks requires $O(n)$ memory for the
rooted tree (e.g., using parent pointers) and an $O(m)$ map for
assigning each worker its position in the tree, totaling $O(n+m)$
memory.

A blockchain tends to achieve synchronization among the workers,
mainly, because of three reasons. First, workers follow a
\textit{standard protocol} in which they are constantly trying to
produce new blocks and broadcasting their achievements to the entire
network. In the case of cryptocurrency networks, for instance, this
behavior is motivated by paying cryptocurrency rewards. Second,
workers can easily verify (i.e., with a fast algorithm) the
authenticity of any block. If a malicious worker changes the
information in one block, it is forced to repeat the extensive
proof-of-work process for that block and all its subsequent blocks in
the blockchain in order for its malicious modification to be made part
of the global blockchain. Since this requires to spend enough
electricity, time, and/or machine rental, such a situation is hardly
ever expected to happen. Third, the standard protocol forces any
malicious worker to confront the computational power of the whole
network, assumed to have mostly honest nodes, into a race that is very
hard to win~\citep{pinzon2016double}.

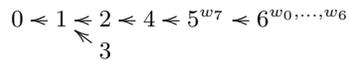
\begin{figure}
  \centering
$\xymatrix{\xyR{0pc}\xyC{0.5pc}0 & 1\ar[l] & 2\ar[l] & 4\ar[l] & 5^{w_{7}}\ar[l] & 6^{w_{0},...,w_{6}}\ar[l]\\
 &  & 3\ar[ul]
}
$
  \caption{\label{fig.pow.blockchainslow}Example of a typical slow system with few
and short branches.}
\end{figure}

Algorithm~\ref{alg.pow.protocol} presents a definition of the
above-mentioned standard protocol assumed to be followed for each
worker in a blockchain system. It can be summarized as follows.  When
a worker produces a new block, it appends a leave to the block it is
standing on, moves to it, and notifies the network about its current
position and new distance to the root. When a worker receives a
notification, it compares its distance to the root with the incoming
position and switches to it whenever the one received is larger.  To
illustrate the use of the standard protocol with a simple example,
consider the blockchains depicted in figures~\ref{fig.pow.blochainrep}
and~\ref{fig.pow.blockchainslow}, respectively. In the former, either
$w_{1}$ or $w_{4}$ produced block 6, but the other workers are not yet
aware of its existence. In the latter, most of the workers are
synchronized with the longest branch, which is typical of a slow
blockchain system, resulting in a tree with few and short branches.

\begin{myAlgorithm}{Standard protocol for each worker $w_{i}$ in a blockchain.}
\label{alg.pow.protocol}

$B_{i}\leftarrow[\text{origin}]$

\textbf{do forever}\algoIndent{\textbf{do~in~parallel,~stop~on~first~to~occur}\algoIndent{Task~1:\textbf{~}$b\leftarrow\text{produce a subsequent block for \ensuremath{B_{i}}}$

Task~2:\textbf{~}$B'\leftarrow\text{notification from another worker}$}

\textbf{end}

\textbf{if~}Task~1~was~completed\textbf{~then}\algoIndent{append~$b$~to~$B_{i}$

$\text{notify workers in the network about \ensuremath{B_{i}}}$}\textbf{else~if~}$B'$\textbf{~}is~longer~than~$B_{i}$\textbf{~then}\algoIndent{$B_{i}\leftarrow B'$}\textbf{endif}}

\end{myAlgorithm}

Some final remarks on inter-node communication, important for
blockchain implementations enforcing the standard protocol, are
due. Note that message communication in the standard protocol is
required to include enough information about the position of a worker
to be located in the tree. The degree of this information depends,
generally, on the system's design itself. On the one hand, sending the
complete sequence from root to end as part of such a message is a
precise way of doing it, but also an expensive one in terms of
bandwidth, computation, and time. On the other hand, sending only the
last block as part of the message is modest on resources, but it could
represent a communication conundrum whenever the worker being notified
about a new block $x$ is not yet aware of $x$'s parent block. This
does not happen very often in slow blockchains, but can be frequent in
fast systems --such as the ones analyzed in a later section of this
manuscript. The common solution, in this case, is to use subsequent
messages to query about the previous blocks of $x$, as needed, thus
extending the average duration of inter-working communication.

\section{A Random Network Model for Blockchains}
\label{sec.model}

The network model proposed in this paper generates a rooted tree
representing a global blockchain from a collection of linked lists
representing local blockchains (see
Definition~\ref{def.pow.blockchain}). It consists of three main
mechanisms, namely, growth, attachment, and broadcast. By growth it is
meant that the number of blocks in the network increases by one at
each step. Attachment refers to the fact that new blocks connect to an
existing block, while broadcast refers to the fact that the new
connected block is announced to the network. In mathematical terms,
the network model is parametric in a natural number $m$ specifying the
number of workers, and two probability distributions $\alpha$ and $\beta$
governing the growth, attachment, and broadcast mechanisms of a
blockchain. Internally, the growth mechanism creates a new block to be
assigned at random among the $m$ workers by taking a sample from
$\alpha$ (the time it took to produce such a block) and broadcasts a
synchronization message to the entire network, whose reception time is
sampled from $\beta$ (the time it takes the other workers in the
network to update their local blockchains with the new block).

A network at a given discrete step $n$ is represented as a rooted tree
$T_n = (V_n, E_n)$, with nodes $V_n \subseteq \nats$ and edges $E_n
\subseteq V_n \times V_n$, and a map $w_n : \{ 0,1,\ldots, m-1 \} \to
V_n$. A node $u \in V_n$ represents a block $u$ in the network and an
edge $(u, v) \in E_n$ represents a directed edge from block $u$ to its
\textit{parent} block $v$. The assignment $w_n(w)$ denotes the
position (i.e., the last block in the local blockchain) of worker $w$
in $T_n$.

\begin{definition}\label{def.model.def}
  Let $\alpha$ and $\beta$ be positive and non-negative probability
  distributions respectively. The algorithm used in the network model
  starts with $V_0=\{b_0\}$, $E_0=\{\}$ and $w_0(w)=b_0$ for all
  workers $w$, being $b_0=0$ the root block, and at each step $n > 0$,
  it evolves as follows:
  \begin{description}
    \item[\textnormal{Growth.}] A new block $b_n$ (or, simply, $n$) is
      created with production time $\alpha_n$ sampled from $\alpha$.
      That is, $V_n = V_{n-1}\cup\{n\}$.
    \item[\textnormal{Attachment.}] Uniformly at random, a worker $w
      \in \{0,1,\ldots,m-1\}$ is chosen for the new block to extend
      its local blockchain. A new edge appears so that $E_n =
      E_{n-1}\cup\{(w_{n-1}(w), n)\}$, and $w_{n-1}$ is updated to
      form $w_n$ with the new assignment $w \mapsto n$, that is,
      $w_n(w) = n$ and $w_n(z)=w_{n-1}(z)$ for any $z \neq w$.
    \item[\textnormal{Broadcast.}] The worker $w$ broadcasts the
      extension of its local blockchain with the new block $n$ to all
      other workers $z$ with times $\beta_{n, z}$ sampled from
      $\beta$.
  \end{description}
\end{definition}

The rooted tree generated by the model in
Definition~\ref{def.model.def} begins with the block $0$ (the root)
and adds new blocks $n=1,2,\ldots$ to some of the workers. At each
step $n>0$, a worker $w$ is selected at random and its local
blockchain, $0 \leftarrow \cdots \leftarrow w_{n-1}(w)$, is extended
to $0 \leftarrow \cdots \leftarrow w_{n-1}(w) \leftarrow n=w_n(w)$.
This results in a concurrent random global behavior, inherent to
distributed blockchain systems, not only because the workers are
chosen randomly due to the proof-of-work scheme, but also because the
communication delay brings some workers out of sync. It is important
to note that the steps $n=0,1,2,\ldots$ in the model are logical time
steps, not to be confused with the sort of time units sampled from the
variables $\alpha$ and $\beta$.  More precisely, the model does not
mention explicitly the time advancement but assumes that workers are
synchronized at the corresponding point in the logical future. For
instance, if $w$ sends a synchronization message of a newly created
block $n$ to another worker $z$, at the end of logical step $n$ and
taking $\beta_{n,z}$ time, then the model assumes that the message
will be received by $z$ during the logical step $n' \geq n$ that
satisfies $\sum_{i=n+1}^{n'}\alpha_{i} \leq
\beta_{n,z} < \sum_{i=n+1}^{n'+1}\alpha_{i} $.

Other reasonable assumptions are implicitly made in the proposed
model, namely: (i) the computational power of all workers is similar;
and (ii) any broadcasting message includes enough information about
the new block and its previous blocks, that no re-transmission is
required to fill block gaps or, equivalently, that these
re-transmission times are included in time sampled from $\beta$.
Assumption (i) justifies why the worker producing the new block is
chosen uniformly at random. Thus, instead of simulating the
proof-of-work of the workers in order to know who will produce the
next block and at what time, it is enough to select a worker uniformly
and to take a sample time from $\alpha$. Assumption (ii) helps in
keeping the model description simple because, otherwise, it would be
mandatory to explicitly define how to proceed when a worker is
severely out of date, that is, if it requires several messages to
synchronize.

In practice, the distribution $\alpha$ that governs the time it takes
for the whole network as a single entity to produce a block is
exponential with mean $\overline{\alpha}$. This is due to the
proof-of-work scheme. Since this scheme is based on finding a nonce
that makes a hashing function fall into a specific set of targets, the
process of producing a block is statistically equivalent to waiting
for a success in a sequence of Bernoulli trials.  Such waiting time
would then correspond --at first-- to a discrete geometric
distribution. However, it can as well be approximated by a continuous
exponential distribution function because the time between trials is
very small compared to the average time between successes (usually
fractions of micro seconds against several seconds or minutes). On the
other hand, the choice of the distribution function $\beta$ that
governs the communication delay, and whose mean is denoted by
$\bar \beta$, can heavily depend on the system under consideration and
its communication details (e.g., hardware, protocol).

\section{Algorithmic Analysis of Blockchain Efficiency}
\label{sec.algo}

This section presents an algorithmic approach to the analysis of
blockchain efficiency. The algorithms presented in this section are
used to estimate the proportion of valid blocks that are produced,
based on the network model introduced in Section~\ref{sec.model}, both
for blockchains with bounded and unbounded number of workers up to a
given number of steps.  In general, although presented in this section
for the specific purpose of measuring blockchain efficiency, these
algorithms can be easily modified to compute other metrics of
interest.

\begin{definition}\label{def.algo.valid}
  Let $T_n = (V_n, E_n)$ be a blockchain obtained from
  Definition~\ref{def.model.def}. The \emph{proportion of valid
  blocks} $p_n$ in $T_n$ is defined as the random variable:
  \begin{align*}
    p_n & = \frac{\max\{\textnormal{dist}(0,u) \mid u\in V_n\}}{|V_n|}.
  \end{align*}
  The \emph{proportion of valid blocks} $p$ produced for a
  blockchain (in the limit) is defined as the random variable:
  \begin{align*}
    p & = \lim_{n \to \infty} p_n.
  \end{align*}

  And their expected values are denoted with $\bar{p_n}$ and
  $\bar{p}$ respectively.
\end{definition}

Note that these random variables are particularly useful to determine
some important properties of blockchains. For instance, the
probability that a newly produced block becomes valid in the long run
is exactly $\bar{p}$, the average rate at which the longest branch
grows can be approximated by $\bar{p}/\bar{\alpha}$, the rate at which
invalid blocks are produced is approximately
$(1-\bar{p})/\bar{\alpha}$, and the expected time for a block
receiving one confirmation $\bar{\alpha}/\bar{p}$. Although $p_n$ and
$p$ are random for any single simulation, their expected values
$\bar{p}_n$ and $\bar{p}$ can be approximated by averaging several
Monte Carlo simulations.

The three algorithms presented in this section, which are sequential
and single threaded, are designed to compute the value of $p_n$ under
the assumption of the standard protocol
(Algorithm~\ref{alg.pow.protocol}). They can be used for computing
$\bar{p}_n$ and, thus, for approximating $\bar{p}$ with large values
of $n$. The first and second algorithms \textit{exactly} compute the
value of $p_n$ for a bounded number of workers. The difference is that
the first one simulates the three mechanisms present in the network
model (i.e., growth, attachment, and broadcast --see
Definition~\ref{def.model.def}), while the second one takes a more
time-efficient approach for computing $p_n$. The third one is a fast
approximation algorithm for $p_n$ in the context of an
\textit{unbounded} number of workers; this algorithm is of special
interest for studying the efficiency of large and fast blockchain
systems because its time complexity does not depend on the number of
workers in the network.

\subsection{Network Simulation with a Priority Queue}
\label{sec.algo.pq}

Algorithm~\ref{alg.algo.pq} simulates the network model with $m$
workers running concurrently under the standard protocol up to $n$
logical steps. This algorithm uses a list $B$ of $m$ block sequences
that reflect the internal blockchains of each worker. The sequences
are initially limited to the origin block $0$ and can be randomly
extended during the simulation. Each iteration of the main loop
consists of four stages: (i) waiting for a new block to be produced,
(ii) simulating the reception of messages during a period of time,
(iii) adding a block to the blockchain of a randomly selected worker,
and (iv) broadcasting the new position of the selected worker in the
shared blockchain to the others in the network. The priority queue
\textit{pq} is used to queue messages for future delivery, thus
simulating the communication delays. Messages have the form
$(t',i,B')$, where $t'$ represents the message future (physical)
arrival time, worker $i$ is the recipient, and the content $B'$
informs about a (non-specified) worker having the sequence of blocks
$B'$. The statements $\alpha()$ and $\beta()$ draw samples from
$\alpha$ and $\beta$, respectively.

\begin{myAlgorithm}{Simulation of $m$ workers using a priority queue.}
\label{alg.algo.pq}

$t\leftarrow0$

\emph{$B\leftarrow[\,[0],[0],...,[0]\,]\quad$($m$ block sequences, 0 is the origin)}

pq $\leftarrow$ empty priority queue

\textbf{for} $k\leftarrow1,...,n-1$ \textbf{do}\algoIndent{$t\leftarrow t+\alpha()$

\textbf{for~}$(t',i,B')\in\text{pq}$~with~$t'<t$~\textbf{~do}\emph{$\quad$(receive~notifications)}\algoIndent{pop~$(t',i,B')$~from~pq

\textbf{if~}$B'$\textbf{~}is~longer~than~$B_{i}$\textbf{~then~}$B_{i}\leftarrow B'$~\textbf{endif}}\textbf{end}

$j\leftarrow\text{random\_worker()}$\emph{$\quad$(producer)}

append~a~new~block~$(k)$~to~$B_{j}$

\textbf{for~}$i\in\left\{ 0,...,m-1\right\} \setminus\left\{ j\right\} $\textbf{~do}~\emph{$\quad$(publish~notifications)}\algoIndent{push~$(t+\beta(),i,B_{j})$~to~pq

}\textbf{end}}

\textbf{end}

$s\leftarrow\underset{s\in B}{\arg\max}\left|s\right|$\emph{$\quad$(longest sequence)}

\textbf{return} $|s|/n$
\end{myAlgorithm}

The overall complexity of Algorithm~\ref{alg.algo.pq} depends, as
usual, on specific assumptions on its concrete implementation. First,
it is assumed that the time complexity to query $\alpha()$ and
$\beta()$ is $O(1)$, which is typical in most computer programming
languages). However, the time complexity estimates presented next may
be higher depending on their specific implementations; e.g., if a
histogram is used instead of a continuous function for sampling these
variables.  If the statement $B_{i}\leftarrow B'$ is implemented
creating a copy in $O(n)$ time and the append statement is $O(1)$,
then the overall time complexity of the algorithm is
$\Omega(mn^{2})$. If $B_{i}\leftarrow B'$ merely copies the list
reference in $O(1)$ and the append statement creates a copy in $O(n)$,
the complexity is improved down to $O(mn\log(mn))$, under the
assumption of having a standard priority queue with log-time insertion
and removal. In either case, the spatial complexity is $O(mn)$.

One key advantage of this algorithm, with respect to the other ones
presented next, is that it can be slightly modified to return the
blockchain $s$ instead of the proportion $p_n$. This would enable a
richer analysis to be carried out in the form of other metrics
different to $p$.  For example, assume $I$ denotes the random variable
that describes the quantity of invalid blocks that are created between
consecutive blocks. $E[I]$ can be estimated from $\bar{p}$ because
$E[I] \approx (1-\bar{p}) / \bar{\alpha}$. However, building a
complete blockchain can be used to estimate not only $E[I]$, but also
a complete histogram of $I$ and all the properties it may posses.

\subsection{A Faster Simulation Algorithm}
\label{sec.algo.faster}

Algorithm~\ref{alg.algo.faster} can be a faster alternative to
Algorithm~\ref{alg.algo.pq}. It uses a different encoding for the
collection of local blockchains: it stores their length instead of the
sequences themselves and suppresses the need for a priority queue.

\begin{myAlgorithm}{Simulation of $m$ workers using a matrix $d$}
\label{alg.algo.faster}

$t_{0},h_{0},z_{0}\leftarrow0,1,0$

$\vec{d_{0}}\leftarrow\left\langle 0,0,...,0\right\rangle \quad$\emph{ (}$m$\emph{ elements)}

\textbf{for} $k\leftarrow1,...,n-1$ \textbf{do}\algoIndent{$j\leftarrow\text{random\_worker()}$

$t_{k}\leftarrow t_{k-1}+\alpha()$

$h_{k}\leftarrow1+\max\left\{ h_{i}\,|\,i<k\,\wedge\,t_{i}+d_{i,j}<t_{k}\right\} \quad$\textrm{(Algorithm~\ref{alg.algo.prune})}

$z_{k}\leftarrow\max(z_{k-1},h_{k})$

$\vec{d_{k}}\leftarrow\underset{\overbrace{j\text{'th position}}}{\left\langle \beta(),...,\beta(),0,\beta(),...,\beta()\right\rangle }$}

\textbf{end}

\textbf{return} $z_{n-1}$
\end{myAlgorithm}

The variable $t_k$ represents the (absolute) time at which the block
$k$ is created, the variable $h_k$ the length of the local blockchain
after being extended with block $k$, and $z_{k}$ the cumulative
maximum given by $z_{k}:=\max\left\{ h_{i} \mid i\leq k\right\}$.

The spatial complexity of Algorithm~\ref{alg.algo.faster} is $O(mn)$
due to the computation of matrix $d$ and its overall time complexity
is $O(nm+n^2)$ when Algorithm~\ref{alg.algo.prune} is not used. This
is because there are $n$ iterations, each requiring $O(n)$ and $O(m)$
for computing $h_k$ and $\vec{d_k}$, respectively. However, if
Algorithm~\ref{alg.algo.prune} is used for computing $h_k$, the
average overall complexity is reduced. Although the complexity of
Algorithm~\ref{alg.algo.prune} is theoretically $O(k)$ in the worst
case, the experimental evaluations suggest an average below
$O(\bar{\beta}/\bar{\alpha})$ (constant respect to $k$). Thus, the
average runtime complexity of Algorithm~\ref{alg.algo.faster} is
actually bounded by $O\left(nm+ \min\{n^2, n + n \bar{\beta} /
\bar{\alpha} \}\right)$, and this corresponds to $O(nm)$, unless the
blockchain system is extremely fast $(\bar{\beta} \gg \bar{\alpha})$.

\begin{myAlgorithm}{Fast computation of $h_{k}$ given $t_i, z_i, h_i$ and $\vec{d_i}$ for all $i<k$}
\label{alg.algo.prune}

$x,i\leftarrow1,k-1$

\textbf{while }$i\geq0$ and\textbf{ $x<z_{i}$ do}\algoIndent{\textbf{if}~$t_{i}\leq t_{k}-d_{i,j}$~and~$h_{i}>x$\textbf{~then}~\algoIndent{$x=h_{i}$}\textbf{~endif}

$i\leftarrow i-1$}

\textbf{end}

\textbf{return} $1+x$$\quad$\emph{(computes }$h_k := 1+\max\left\{ h_{i}\,|\,i<k\,\wedge\,t_{i}+d_{i,j}<t_{k}\right\} \cup\left\{ 1\right\} $\emph{)}

\end{myAlgorithm}

\subsection{An Approximation Algorithm for Unbounded Number of Workers}
\label{sec.algo.infty}

Algorithms~\ref{alg.algo.pq} and~\ref{alg.algo.faster} are proposed to
compute the value of $p_n$ for a \textit{fixed} number $m$ of workers
in the blockchain. Of course, these algorithms can be used to compute
$p_n$ for different values of $m$. However, the time complexity of
these two algorithms heavily depends on the value of $m$, which
presents a practical limitation when faced with the task of analyzing
large blockchain systems. This section presents an algorithm for
approximating $p_n$ for an unbounded number of workers and
formal observations that support this claim.

Recall the definition of $p_n$, from the beginning of this section,
used as a measure of efficiency in terms of the proportion of valid
blocks in the blockchain $T_n = (V_n, E_n)$ produced up to step $n$:
\begin{align*}
  p_n & = \frac{\max\{\textnormal{dist}(0,u) \mid u\in V_n\}}{|V_n|}.
\end{align*}
This definition assumes a fixed number $m$ of workers. That is $p_n$,
can be better written as $p_{m,n}$ to represent the proportion of
valid blocks in the blockchain $T_n$ \textit{with} $m$ workers. For
the analysis of large blockchains, the challenge is then in finding an
efficient way to estimate $p_{m,n}$ as $m$ and $n$ grow. To be more
precise, the challenge is in finding an efficient algorithm for
approximating the random variables $p^*_n$ and $p^*$ defined as
follows:
\begin{align*}
  p^*_n & =  \lim_{m \to \infty} p_{m,n} &&\textrm{and}& p^* & =  \lim_{n \to \infty} p^*_n.
\end{align*}
The approach presented next to tackle the given challenge is to modify
Algorithm~\ref{alg.algo.faster} by removing the matrix $d$.  The idea
is to replace the need for the $d_{i,j}$ values by an approximation
based on the random variable $\beta$ in order to compute $h_k$ in each
iteration of the main loop. The first observation is that the first
row can be assumed to be 0 wherever it appears because $d_{0,j}=0$ for
all $j$. For the remaining rows, an approximation is introduced by
observing that if an element $X_{m}$ is chosen at random from the
matrix $d$ of size $(n-1)\times m$ (i.e., matrix $d$ without the first
row), the cumulative distribution function of $X_{m}$ is given by
\[
P(X_{m}\leq r)=\begin{cases}
0 & ,\; r<0\\
\frac{1}{m}+\frac{m-1}{m}P(\beta() \leq r) &,\; r\geq0,
\end{cases}
\]

\noindent where $\beta()$ is a sample from $\beta$. This is because
the elements $X_{m}$ of $d$ are either samples from $\beta$, whose
domain is $\mathbb{R}_{\geq 0}$, or the value $0$ with a probability
of $1/m$ since there is one zero per row. Therefore, given that the
following functional limit converges uniformly
(Theorem~\ref{thm.algo.conv}),
\[
\lim_{m\to\infty}\left(r \overset{f_m}{\mapsto} P(X_{m}\leq r)\right)\overset{\text{}}{=}\left(r \overset{f}{\mapsto} P(\beta()\leq r)\right),
\]
\noindent each $d_{i,j}$ can be approximated by directly sampling the
distribution $\beta$ and then Algorithm~\ref{alg.algo.prune} can be
used for computing $h_k$ by replacing $d_{i,j}$ with $\beta()$.

\begin{theorem}\label{thm.algo.conv}
  Let $f_{k}(r):=P(X_{k}\leq r)$ and $g(r):=P(\beta()\leq r)$.  The
  functional sequence $\left\{ f_{k}\right\} _{k=1}^{\infty}$
  converges uniformly to $g$.
\end{theorem}

\begin{proof}
Let $\epsilon>0$. Define $n:=\left\lceil \frac{1}{2\epsilon}\right\rceil $
and let $k$ be any integer with $k>n$. Then
\begin{align*}
  \sup\left|f_{k}-g\right|= & \sup\left\{ \left|\frac{1}{k}+\left(\frac{k-1}{k}-1\right)P(\beta()\leq r)\right|:r\geq0\right\} \\
  \leq & \frac{1}{k}+\frac{1}{k}\sup\left\{ P(\beta()\leq r):r\geq0\right\} \\
  = & \frac{1}{k}+\frac{1}{k}\\
  < & \frac{2}{n}\leq\epsilon.
\end{align*}\qed
\end{proof}

With the help of the above observations, the need for the bookkeeping
matrix is removed, and both variables $j$ and $d$ can be discarded
from Algorithm~\ref{alg.algo.faster} to obtain
Algorithm~\ref{alg.algo.infty}. Note that this new algorithm computes
$p^*_n$, an approximation of $\lim_{m\to\infty} p_{m,n}$ in which the
matrix entries $d_{i,j}$ are replaced by new samples from $\beta$ each
time they are needed, thus ignoring the arguably negligible hysteresis
effects.

\begin{myAlgorithm}{Approximation for
$\lim_{m\to\infty}p_{m,n}$ simulation}
\label{alg.algo.infty}

$t_{0},h_{0},z_{0}\leftarrow0,0,0$

\textbf{for} $k\leftarrow1,...,n-1$ \textbf{do}
\algoIndent{
$t_{k}\leftarrow t_{k-1}+\alpha()$

$h_{k}\leftarrow1+\max\left\{ h_{i} \mid i<k\,\wedge\,t_{i}+\beta()<t_{k}\right\} \cup\left\{ 1\right\}$ \qquad \textrm{(Algorithm~\ref{alg.algo.prune}*)}

$z_{k}\leftarrow\max(z_{k-1},h_{k})$}

\textbf{end}

\textbf{return} $z_{n-1}$

{\nolinenumber \textrm{Algorithm~\ref{alg.algo.prune}* stands for
Algorithm~\ref{alg.algo.prune} with $\beta()$ instead of $d_{i,j}$
(approximation)}}
\end{myAlgorithm}

The complexity of Algorithm~\ref{alg.algo.infty}, if implemented
without using Algorithm~\ref{alg.algo.prune}, is $O(n^{2})$-time and
$O(n)$-space. But if the prunning algorithm is used, the time
complexity drops below $O(n + n \bar\beta / \bar\alpha))$ according to
experimentation, which can be considered $O(n)$ as long as the
blockchain system is not extremely fast (i.e., when $\bar{\beta} \gg
\bar{\alpha}$).

\section{Empirical Evaluation of Blockchain Efficiency}
\label{sec.results}

This section presents an experimental evaluation of blockchain
efficiency in terms of the proportion of valid blocks produced by the
workers for the global blockchain. The proposed model in
Section~\ref{sec.model} is used as the mathematical basis for the
evaluation, while the algorithms in Section~\ref{sec.algo} are used
for the experimental evaluation.  The main claim made in this section
is that the efficiency of a blockchain, under some assumptions later
identified in this section, can be expressed as a ratio between
$\bar{\alpha}$ and $\bar{\beta}$.  This section also presents
experimental evidence on why Algorithm~\ref{alg.algo.infty} --a fast
approximation algorithm for computing the proportion of valid blocks
in a blockchain with an unbounded number of workers-- is an accurate
tool for computing $p^* = \lim_{n,m\to\infty}p_{m,n}$.

The speed of a blockchain can be characterized by the relationship
between the expected values of $\alpha$ and $\beta$.

\begin{definition}\label{def.results.types}
  Let $\alpha$ and $\beta$ be the distributions in
  Definition~\ref{def.model.def}. A blockchain is called:
  \begin{itemize}
  \item \emph{slow} whenever $\bar{\alpha}\gg\bar{\beta}$.
  \item \emph{chaotic} whenever $\bar{\alpha} \ll \bar{\beta}$.
  \item \emph{fast} whenever $\bar{\alpha} \approx \bar{\beta}$.
  \end{itemize}
\end{definition}

The classification in Definition~\ref{def.results.types} captures the
intuition about the behavior of a global blockchain in terms of how
alike the times for producing a block and for local block
synchronization are. The Bitcoin implementation can be classified as a
slow blockchain because the time between the creation of two
consecutive blocks is much larger than the time it takes the local
blockchains to synchronize. In chaotic blockchains, a dwarfing
synchronization time means that basically no or very little
synchronization is possible, resulting in a blockchain in which rarely
any block would be part of ``the'' valid chain of blocks.  A fast
blockchain, however, is one in which both the times for producing a
block and message broadcasting are similar. The two-fold goal of this
section can now be better explained. First, it is to analyze the
behavior of $\bar{p}^*$ for any of these three classes of
blockchains. On the other, it is to understand how the trade-off
between block production speed and broadcasting time affect the
efficiency of the data structure by means of a formula.

In favor of readability, the experiments presented in this section
identify algorithms~\ref{alg.algo.faster} and~\ref{alg.algo.infty} as
$A_m(\_)$ and $A_\infty(\_)$, respectively. Furthermore, the claims
and experiments presented next assume that the distribution
$\alpha$ is exponential, which is the case for proof-of-work systems.

\begin{myclaim}\label{cla.results.algos}
  Unless the system is chaotic, the hysteresis effect of the matrix
  entries $d_{i,j}$ in Algorithm~\ref{alg.algo.faster} is negligible.
  Moreover, $\lim_{m,n\to\infty}A_m(n) = A_\infty(n)$.
\end{myclaim}

Note that Theorem~\ref{thm.algo.conv} implies that if the hysteresis
effect of the random variables $d_{i,j}$ is ignored, then it would
necessarily hold that $\lim_{m,n\to\infty}A_m(n) = A_\infty(n)$.
However, it does not prove that the equation holds in general, but
only if the hysteresis effect is negligible. Experimental evaluation
suggests that this is the case indeed, as stated in
Claim~\ref{cla.results.algos}.

\begin{figure}
  \noindent \subfloat[Evolution of $A_{m}$ to $A_{\infty}$ as $m$
    grows. At least $100$ samples per point.]  {\noindent
    \includegraphics[width=0.48\textwidth]{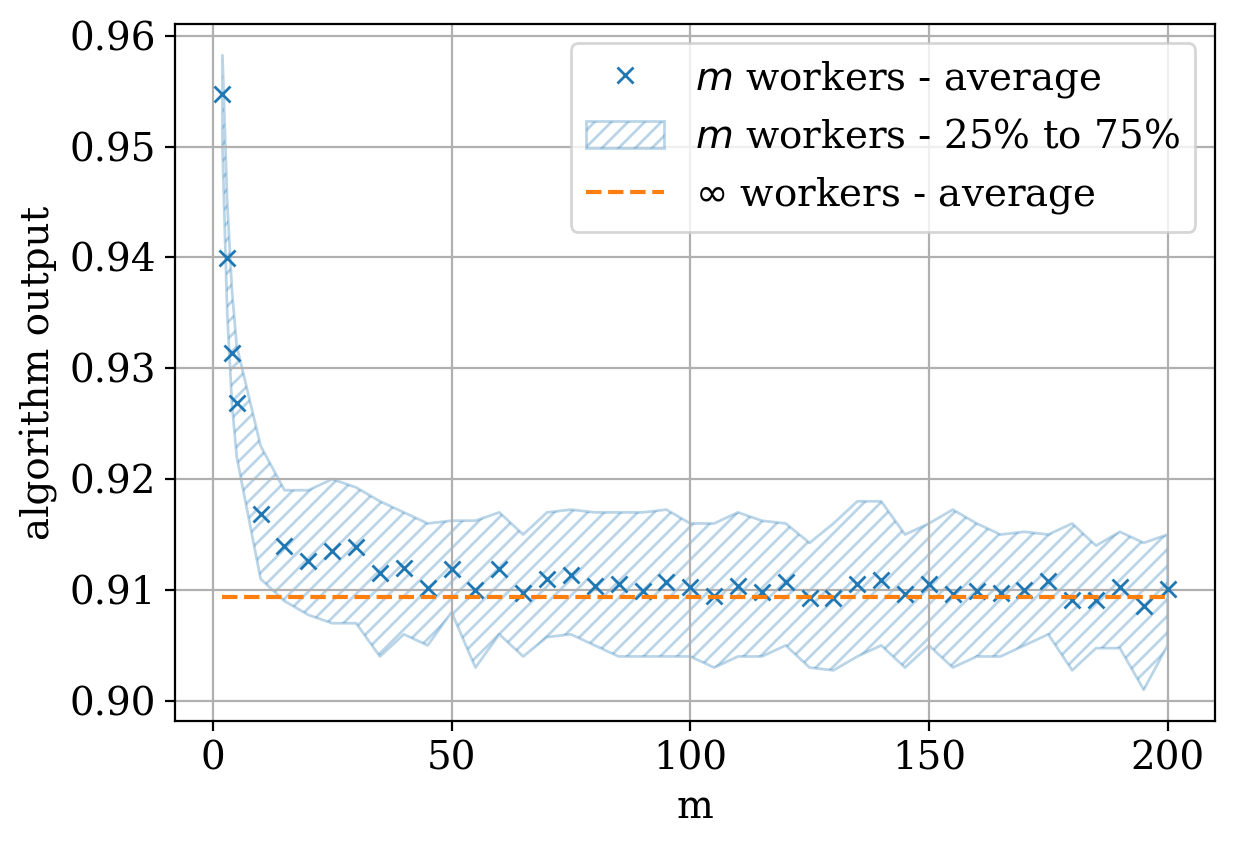}}\hfill{}\subfloat[High
    similarity between the pdf's of $A_{100}$ and $A_{\infty}$. At
    least $1000$ samples in
    total.]{\includegraphics[width=0.48\textwidth]{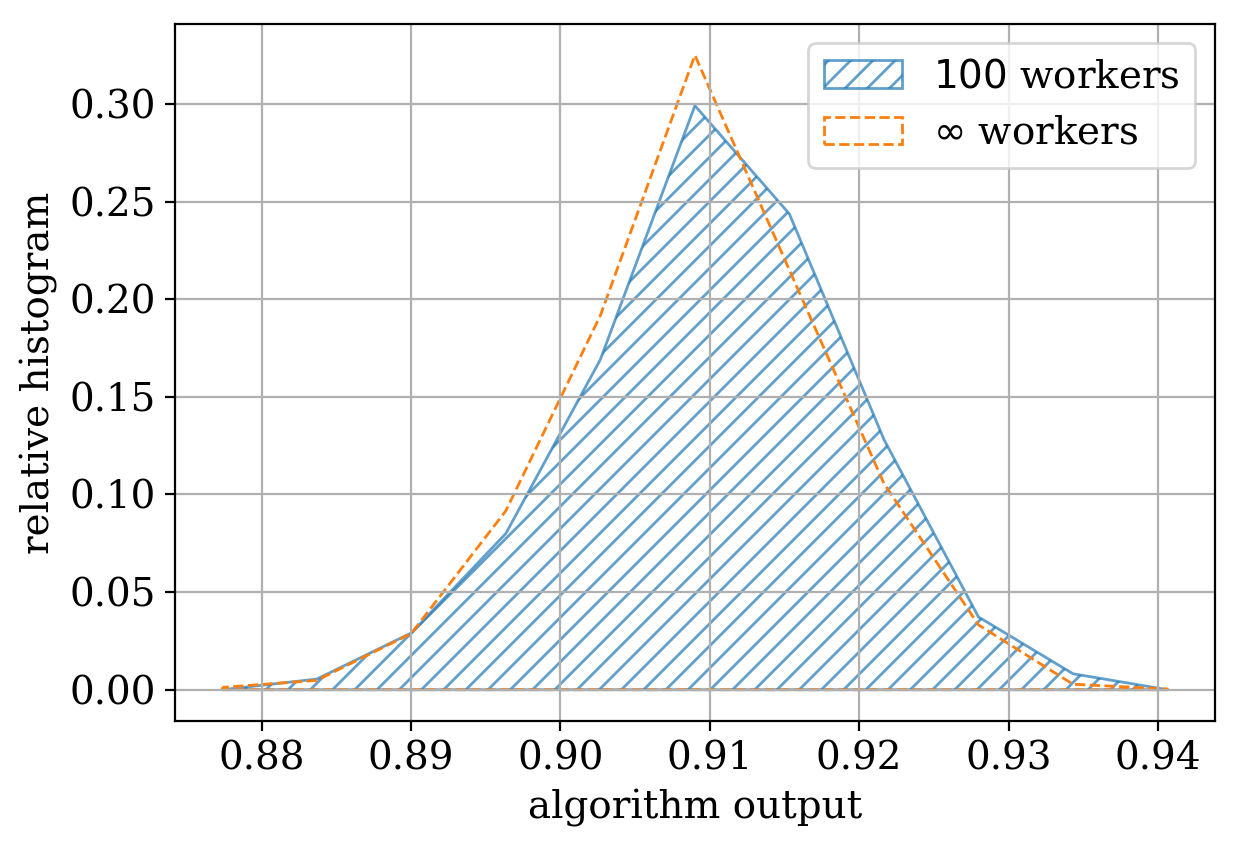}}\caption{\label{fig.results.convergence1}
    Algorithmic simulation of $n=1000$ blocks with
    $\overline{\alpha}=1$, $\overline{\beta}=0.1$, and $\beta$
    exponential. }
\end{figure}

Figure~\ref{fig.results.convergence1} summarizes the average output of
the $A_m$ algorithm and the region that contains half of these
outputs, for several values of $m$. The output of this algorithm seems
to approach that of $A_\infty$, not only for the expected value
(Figure~\ref{fig.results.convergence1}.(a)), but also in p.d.f. shape
(Figure~\ref{fig.results.convergence1}.(b)). These experiments were
performed with several distribution functions for $\beta$ with similar
results. In particular, the exponential, chi-squared, and gamma (with
$k\in\left\{ 1,1.5,2,3,5,10\right\}$) probability distribution
functions were used, all with different mean values, and the plots
were similar to the ones already depicted in
Figure~\ref{fig.results.convergence1}.

\begin{figure}
  \noindent
  \includegraphics[width=0.48\textwidth]{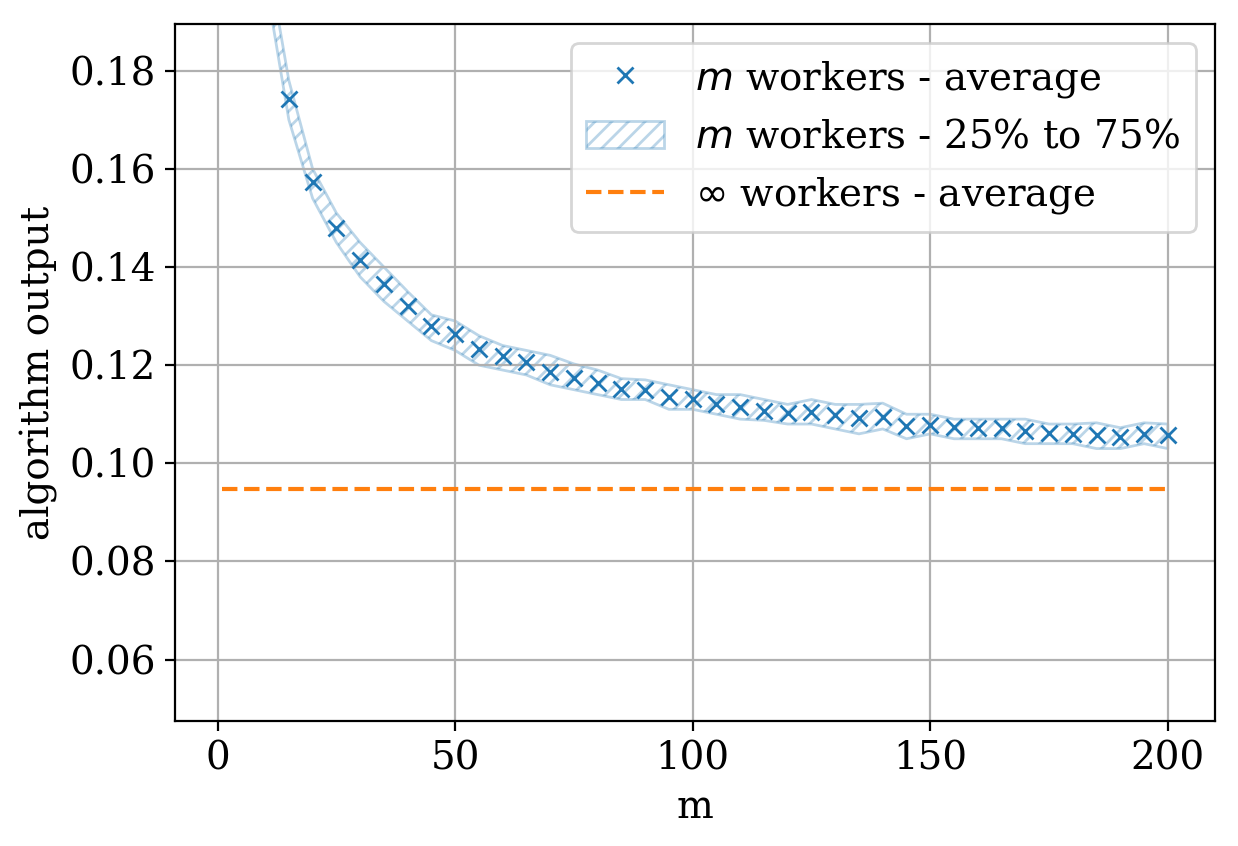}\hfill{}\includegraphics[width=0.48\textwidth]{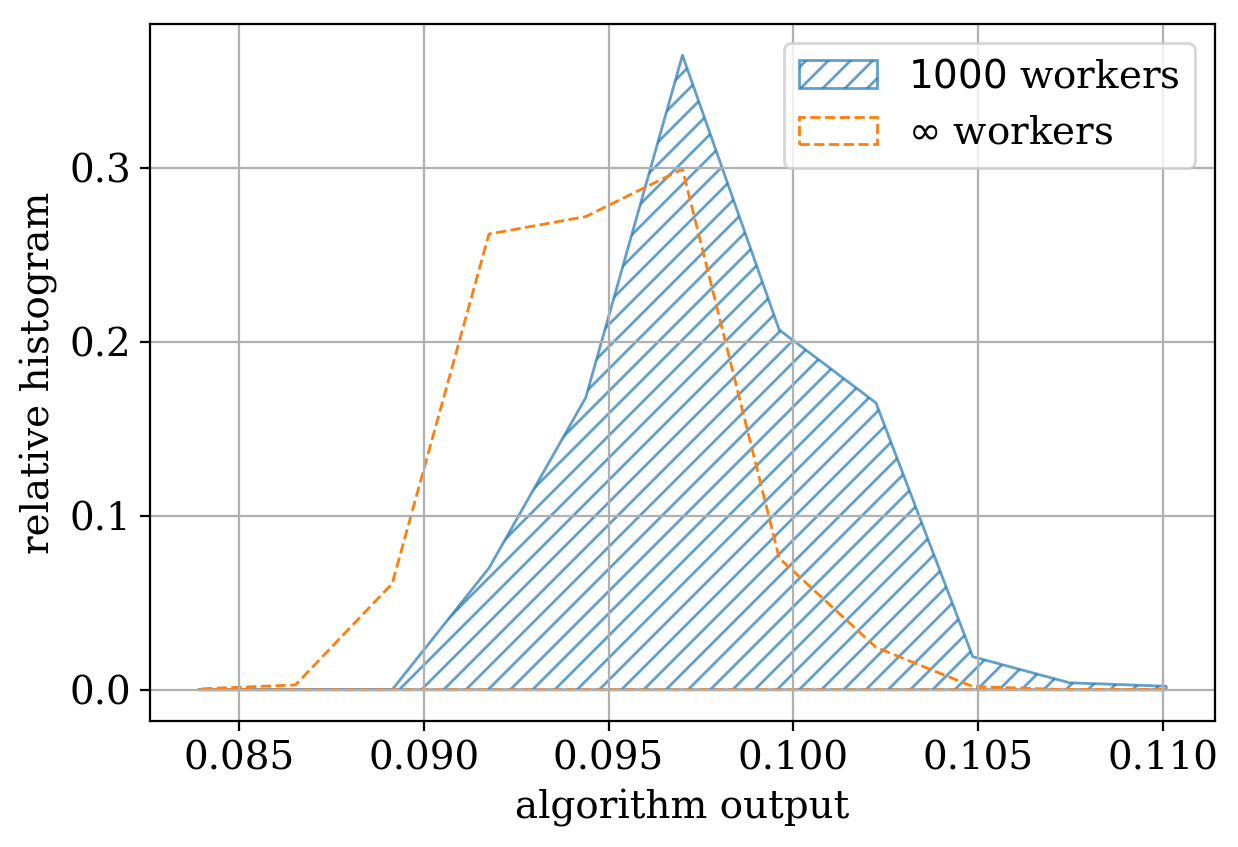}\caption{\label{fig.results.convergence2}The
    convergence is slower and the approximation error is larger in
    chaotic systems: with $1000$ workers there is still an average
    output shift of around $0.005$.}
\end{figure}

However, as the quotient $\overline{\beta}/\overline{\alpha}$ grows
beyond values of $1$, the convergence of $A_m$ becomes much slower and
the approximation error can be noticeable. For instance, this is the
case of a blockchain system in which $10$ blocks are produced in
average during the transmission of a synchronization message (i.e., a
somewhat chaotic system), as depicted in
Figure~\ref{fig.results.convergence2}. Even after considering $1000$
workers, the p.d.f.'s are shifted. The error can be due to the
hysteresis effect that is ignored by the $A_\infty$ algorithm, or it
might be a consequence of the slow rate of convergence. In any case,
the output of these systems is very low, thus making them unstable and
useless in practice.

An intuitive conclusion about blockchain efficiency and speed of block
production is that slower systems tend to be more efficient than
faster ones. That is, faster blockchain systems have a tendency to
overproduce blocks that will not be valid.

\begin{myclaim}\label{cla.results.expected}
  If the system is not chaotic, then \[
  \bar{p}^* =\frac{\bar{\alpha}}{\bar{\alpha}+\bar{\beta}}.
  \]
\end{myclaim}

\begin{figure}
  \includegraphics[width=0.45\textwidth]{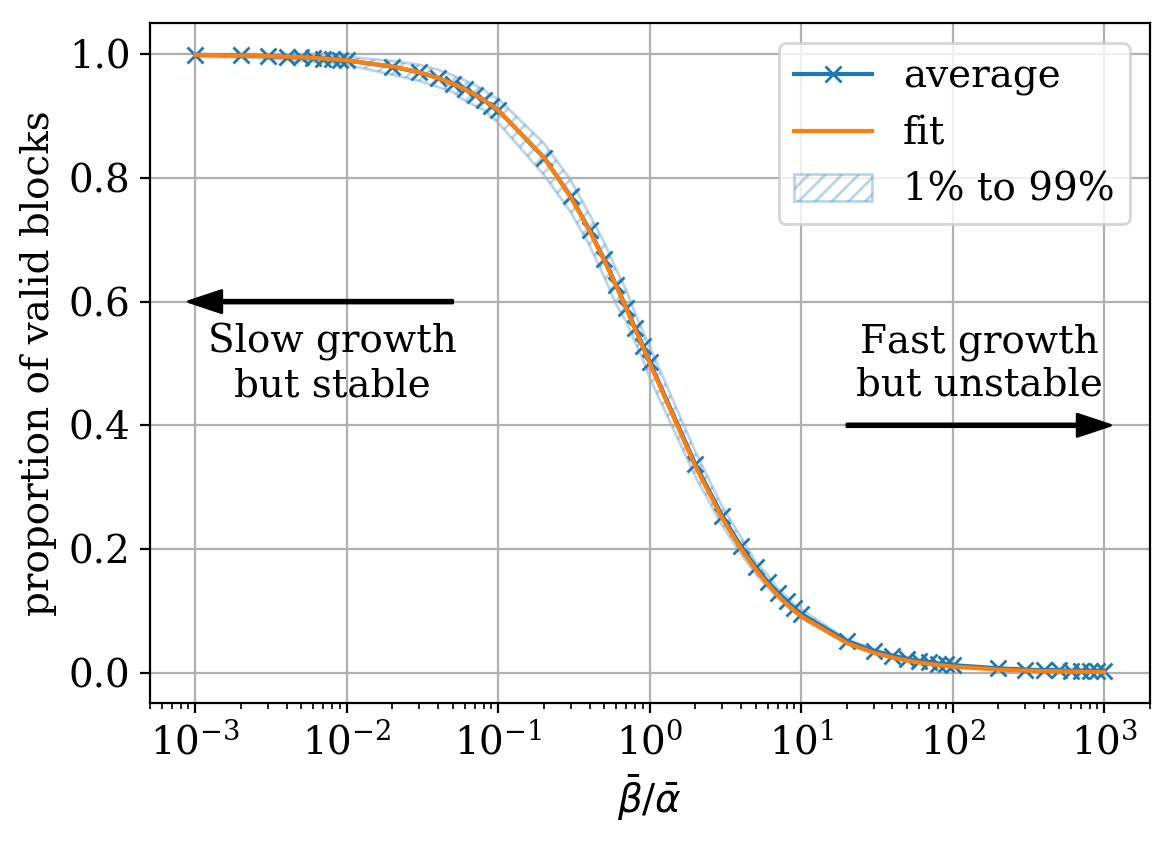}\includegraphics[width=0.45\textwidth]{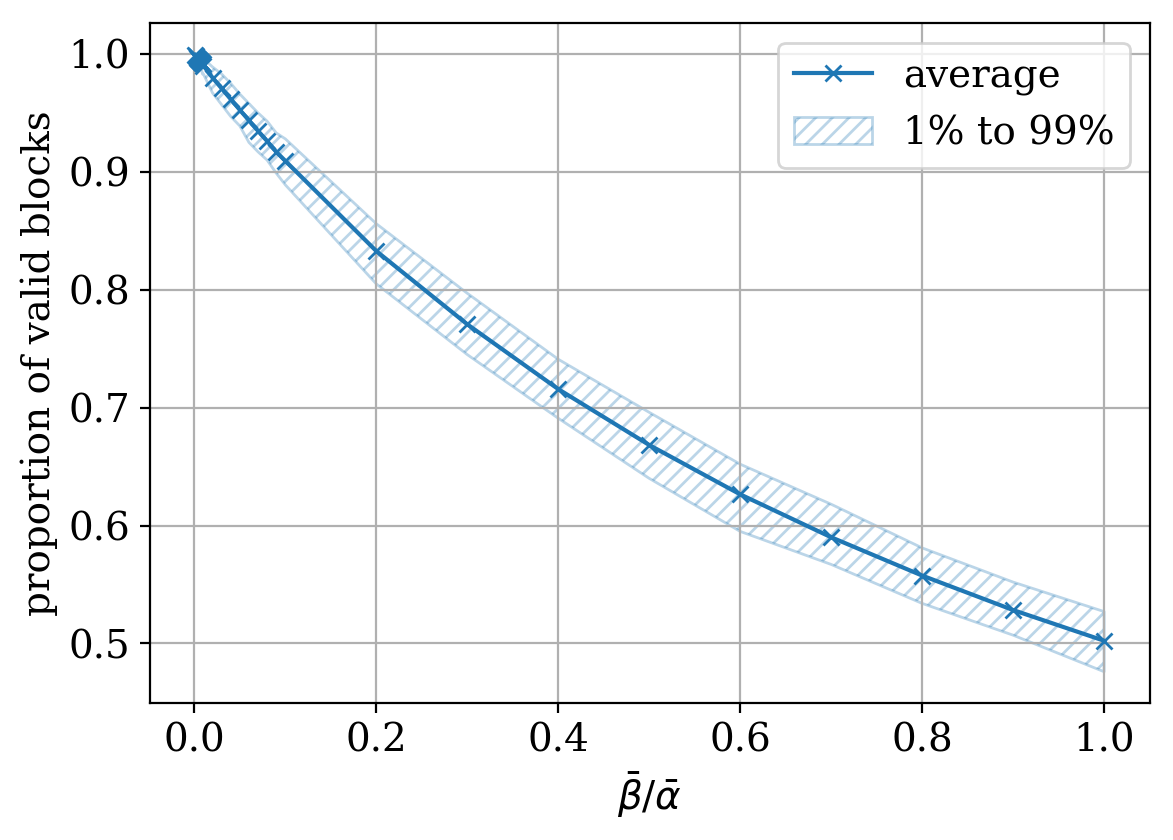}\caption{\label{fig.results.efficiency}Effect
    of speed in the proportion of valid blocks.}
\end{figure}

Figure~\ref{fig.results.efficiency} presents an experimental
evaluation of the proportion of valid blocks produced in a blockchain
in terms of the ratio $\frac{\bar{\beta}}{\bar{\alpha}}$. In both
plots, the horizontal axis represents how fast blocks are produced in
comparison with how slow synchronization is achieved.  If the system
is slow, then efficiency is high because most of valid produced blocks
tend to be valid. If the system is chaotic, however, then efficiency
is low because the newly produced blocks are highly likely to become
invalid. Finally, note that for non-slow blockchains, say for $10^{-1}
\leq \frac{\bar{\beta}}{\bar{\alpha}}$, block production can be good
whenever $\bar{\alpha} > \bar{\beta}$. As a matter of fact, this is an
important observation since the experiments suggest that even if
synchronization of local blockchains takes in average a tenth of the
time it takes to produce a block in general, the proportion of
produced blocks that become valid is near $90\%$. In practice, this
observation can bridge the gap between the current use of blockchains
as slow systems and the need for faster blockchains to cope with the
challenges in the years to come.

\section{Concluding Remarks}
\label{sec.concl}

This paper presented a network model for blockchains and has shown how
the proposed simulation algorithms can be used to analyze the
efficiency (in terms of production of valid blocks) of fast blockchain
systems. The model is parametric on: (i) the number of workers (or
nodes); and (ii) two probability distributions governing the time it
takes to produce a new block and the time it takes the workers to
synchronize their local copies of the blockchain. The simulation
algorithms are probabilistic in nature and can be used to compute the
expected value of several metrics of interest, both for a fixed and
unbounded number of workers, via Monte Carlo simulations.  It is
proven, under some reasonable assumptions, that the fast approximation
algorithm for an unbounded number of workers yields accurate estimates
in relation to the other two exact (but much slower) algorithms.
Claims --supported by extensive experimentation-- have been proposed,
including a formula to measure the proportion of valid blocks produced
in a blockchain in terms of the two probability distributions of the
model. The model, algorithms, experiments, and experimentally backed
insights contributed by this paper can be seen as useful mathematical
tools for specifying, simulating, and analyzing the design of fast
blockchain systems in the broader picture of the years to come.

Future work on the analytic analysis of the experimental observations
contributed in this work should be pursued; this includes proving the
claims. Furthermore, the generalization of the claims to non
proof-of-work schemes, i.e. to different probability distribution
functions for specifying the time it takes to produce a new block may
also be considered. Finally, the study of different forms of attack on
blockchain systems can be pursued with the help of the proposed model.


\bibliographystyle{plain}
\bibliography{refs}

\end{document}